\documentclass[twocolumn,english,showpacs,floatfix,aps,prl]{revtex4}
\usepackage[T1]{fontenc}
\usepackage[latin1]{inputenc}
\usepackage{amsmath}
\usepackage{graphicx}
\usepackage{amssymb}

\makeatletter
\usepackage{graphicx} 
\usepackage{dcolumn} 
\usepackage{bm} 

\usepackage{babel}
\makeatother
\begin{document}

\title{Genuine Multipartite Entanglement in Quantum Phase Transitions}
\tighten

\author{Thiago R. de Oliveira}

\email{tro@ifi.unicamp.br}

\author{Gustavo Rigolin}

\email{rigolin@ifi.unicamp.br}

\author{Marcos C. de Oliveira}

\email{marcos@ifi.unicamp.br}

\affiliation{Instituto de F\'{\i}sica Gleb Wataghin, Universidade
Estadual de Campinas, 13083-970, Campinas-SP, Brazil}

\begin{abstract}
We demonstrate that the Global Entanglement (GE) measure defined
by Meyer and Wallach, J. Math. Phys. \textbf{43}, 4273 (2002), is
maximal at the critical point for the Ising chain in a transverse
magnetic field. Our analysis is based on the equivalence of GE to
the averaged linear entropy, allowing the understanding of
multipartite entanglement (ME) features through a generalization
of GE for bipartite blocks of qubits.  Moreover, in contrast to
GE, the proposed ME measure can distinguish three paradigmatic
entangled states: $GHZ_{N}$, $W_{N}$, and $EPR^{\otimes N/2}$. As
such the generalized measure can detect genuine ME and is maximal
at the critical point.
\end{abstract}


\maketitle


Entanglement is a correlation of exclusively quantum nature
present (in principle) in any set of post-interacting quantum
systems \cite{despagnat}. As such multipartite entanglement (ME)
is expected to play a key role on quantum phase transition (QPT)
phenomena in the same way that (statistical) classical correlation
does on classical phase transitions \cite{Nature,nielsen}.  In
ordinary phase transitions, at the critical point, a non-zero
order parameter characterizes a long range correlation (given by
the correlation length divergence). In the same way, in QPTs it is
expected that ME be maximal at the critical point, in the sense
that all the system parties would be entangled to each other
\cite{nielsen}. However, this conjecture could not be proved in
general neither by measures of pairwise entanglement nor by the
proposed ME measures. Even after a considerable effort, a deep
understanding of multipartite entangled states (MES) is lacked. It
is still a great challenge thus to capture the essential features
of genuine ME, from a conceptual point of view, as well as from a
quantitative approach, defining a measure that among other
properties be able to distinguish MES \cite{cirac,rigolintele}.

Indeed, concerning the legitimate quantum correlations in QPTs it
would be certainly important to know exactly what kind of
entanglement should we expect to be maximal at the critical point.
The great majority of efforts trying to answer this question made
use of two kinds of bipartite entanglement measures, both
calculated for spin-$1/2$ lattice models such as the Ising model
in a transverse magnetic field \cite{Ising
original}. The first one, namely the pairwise entanglement
(concurrence) between two spins in the chain, was studied by Refs.
\cite{Nature,nielsen}. The second one, the entropy of entanglement
between one part of the chain (a block of $L$ spins) and the rest
of the chain, was investigate by Refs.
\cite{nielsen,Latorre,latorre2}. Some candidates of ME measures
were also evaluated in systems exhibiting QPTs \cite{Multipartite
entan,somma,LE}. Nevertheless, none of the entanglement measures
employed in the above references are maximal at the critical point
but the single site entropy for the Ising model \cite{nielsen} in
the thermodynamical limit and the Localizable Entanglement
\cite{LE} for an Ising chain with 14 spins. Furthermore, in Refs.
\cite{nielsen,Nature} the authors have independently shown that
bipartite entanglement vanishes when the distance between the two
spins is greater than two lattice sites \cite{footnote2}. This is
not expected since long range quantum correlations should be
present at the critical point. It was then suggested that
bipartite entanglement at the critical point would be decreased in
order to increase ME due to entanglement sharing \cite{nielsen}.
In other words, ME only appears at the expense of pairwise
entanglement and at the critical point we should expect a genuine
MES.

In this paper we demonstrate that the Global Entanglement (GE)
introduced in Ref. \cite{meyer} indeed captures the essential
point to be maximal at the critical point for the Ising model in a
transverse magnetic field in the thermodynamical limit. We also
prove that there exists an interesting relation among GE, von
Neumann entropy, linear entropy (LE), and $2$-tangle
\cite{coffman,wong,indianos}, showing that they are all equivalent
to detect QPTs. Furthermore, this relation helps us to understand
the results obtained in Ref. \cite{nielsen}, as outlined in the
previous paragraph, and suggests that they are not particular to
the Ising model but common to all MES with translational
invariance. In addition to this, we generalize GE and propose a
new ME measure, which is also maximal at the critical point for
the Ising model, can detect genuine MES, and contrary to GE,
furnishes different values for the entanglement of the $GHZ_{N}$,
$W_{N}$, and $EPR^{\otimes N/2}$ states, thus being able to
distinguish among MES.

For a N qubit system (spin-1/2 chain) it was noticed that GE is
simply related to the $N$ single qubit purities
\cite{brennen,indianos,viola} by
\begin{equation}\vspace{-.2cm}
E_{G}^{(1)}=2-\frac{2}{N}\sum_{j=1}^{N}\text{Tr}(\rho_{j}^{2})
=\frac{1}{N}\sum_{j=1}^{N}S_{L}(\rho_{j})=\langle S_{L}\rangle,
\label{important2}
\end{equation}
 where GE is here on identified as $E_{G}^{(1)}$, $\rho_{j}= \text{Tr}_{\bar j}\{\rho\}$
  is the $j$-th qubit reduced density
 matrix obtained by tracing out the other $\bar j$ qubits,
and
$S_{L}(\rho_j)=\frac{d}{d-1}\left[1-\text{Tr}\left(\rho_{j}^{2}\right)\right]$
is the standard definition of LE. This relation shows that
$E_{G}^{(1)}$ is just the mean of LE. It was also noticed in Refs.
\cite{indianos,endrejat} that
 \begin{equation}\vspace{-.2cm}
E_{G}^{(1)}=\frac{1}{N}\sum_{j=1}^{N}\tau_{j,\,
rest}=\langle\tau\rangle,\label{important3}
\end{equation} where $\tau_{j,\, rest}=C^{2}$ is the
2-tangle \cite{coffman,wong,indianos}, the
 square of the concurrence $C$ \cite{wootters}. Both LE and
 the 2-tangle can thus be used  to quantify the entanglement between any
block bipartition of a system of N-qubits. (They quantify the
entanglement between one qubit $j$ and the {\it rest} $N-1$ qubits
of the chain \cite{indianos}.) The proof of (\ref{important3}) is
based on the Schmidt decomposition \cite{Schmidt}, which also
allows us to use for pure systems the reduced von Neumann entropy,
$ S_{V}(\rho_{j(\bar{j})}) =  -\text{Tr}_{j(\bar
j)}\left[\rho_{j(\bar{j})}\log_{d}(\rho_{j(\bar{j})})\right]$,
 as a good bipartite entanglement measure \cite{Ben96}. Here $d=\min\{
\dim\mathcal{H}_{j},\dim\mathcal{H}_{\bar j}\}$ and
$\dim\mathcal{H}_{j(\bar j)}$ is the Hilbert space dimension of
subsystem $j(\bar j)$. Recalling that $S_{V}$ is bounded from
below by $S_{L}$ and employing Eqs.~(\ref{important2}) and
(\ref{important3}) we obtain the following important relation
\begin{equation}
E_{G}^{(1)}=\langle\tau\rangle=\langle S_{L}\rangle\leq\langle
S_{V}\rangle, \label{importantfinal}
\end{equation}
which states that GE is nothing but the mean LE of single qubits
with the rest of the chain. Furthermore, GE is also equal to the
mean 2-tangle and a lower bound for the mean von Neumann entropy.
An immediate consequence of this result shows up when we deal with
linear chains with translational invariance. This implies that
$\langle S_{L}\rangle=S_{L}(\rho_{j})$ and that $\langle
S_{V}\rangle=S_{V}(\rho_{j})$. Hence, Eq.~(\ref{importantfinal})
becomes $E_{G}^{(1)}=S_{L}(\rho_{j})\leq S_{V}(\rho_{j})$. Since
$S_{L}(\rho_{j})$ and $S_{V}(\rho_{j})$ have the same concavity
and both entropies attain their maximal value for a maximally
mixed state
this last relation shows that $E_{G}^{(1)}$ is as efficient as the
linear and the von Neumann entropies to detect QPTs. In Ref.
\cite{nielsen} the authors used $S_{V}$ and in Ref. \cite{somma}
$E_{G}^{(1)}$ was employed to detect QPTs in the Ising model.
Needless to say, both works arrived at the same results for a
given range of parameters via, notwithstanding, different
entanglement measures which by that time were thought to be
unrelated.

Despite its success to detect the Greenberger-Horne-Zeilinger
(GHZ) state \cite{ghz,endrejat}, $E_{G}^{(1)}$ sometimes fails for
distinguishing different multipartite states. This is best
understood if we study $E_{G}^{(1)}$ for three paradigmatic
multipartite states.
The first is 
$|GHZ_{N}\rangle=(1/\sqrt{2})
\left(|0\rangle^{\otimes N}+|1\rangle^{\otimes N}\right)$,
 where $|0\rangle^{\otimes N}$ and $|1\rangle^{\otimes N}$ represent
$N$ tensor products of 
$|0\rangle$ and $|1\rangle$
respectively. 
The second 
is 
a tensor product of $N/2$ Bell states \cite{viola},
$|EPR_{N}\rangle=|\Phi^{+}\rangle^{\otimes \frac{N}{2}}$,
 where $|\Phi^{+}\rangle=(1/\sqrt{2})(|00\rangle+|11\rangle)$. 
This state is obviously not a MES. Only the pairs of qubits
$(2j-1,2j)$, where $j=1,2,...,N$, are entangled. Nevertheless, for
both states $E_{G}^{(1)}=1$. 
The last one 
is the W state \cite{cirac}: 
$|W_{N}\rangle=(1/\sqrt{N})\sum_{j=1}^{N}|00\cdots1_{j}\cdots00\rangle$.
 The state $|00\cdots1_{j}\cdots00\rangle$ represents 
 $N$ qubits 
 in which the $j$-th 
 is $|1\rangle$ and 
 the others are $|0\rangle$. As shown in Ref. \cite{meyer},
$E_{G}^{(1)}(W_{N})=4(N-1)/N^{2}$. 

We now present a generalization of GE. The main features of this
new approach are three-fold. First, it 
becomes clear that we 
have different classes of ME measures, where $E_{G}^{(1)}$ is the
first one. Second, the first non trivial class, $E_{G}^{(2)}$,
furnishes different values for the three states considered above 
Third, it gives new insights in the study of QPT and ME.

In order to define $E_{G}^{(2)}$ we need the following 
function
\begin{equation}
G(2,l)\equiv\frac{4}{3}\left(1-\frac{1}{N-l}\sum_{j=1}^{N-\;
l}\text{Tr} \left(\rho_{j,j+l}^{2}\right)\right),
\label{funcaoG}
\end{equation}
where $\rho_{j,j+l}$ is the 
density matrix of 
qubits $j$ and $j+l$, 
obtained by tracing out the other
$N-2$ qubits. The index $0<l<N$
is the distance in the chain of two qubits 
and 
$4/3$ is a normalization constant 
assuring $G(2,l)\leq 1$. 
Of interest here are two quantities that can
be considered 
ME measures in the same sense that $E_{G}^{(1)}$ is: 
\vspace{-.2cm}
\begin{eqnarray}
G(2,1) &\equiv& \frac{4}{3}\left(1-\frac{1}{N-1}\sum_{j=1}^{N-\;1}
\text{Tr}\left(\rho_{j,j+1}^{2}\right)\right),\label{g21}\end{eqnarray}
\vspace{-.5cm}and\begin{eqnarray} E_{G}^{(2)} & = &
\frac{1}{N-1}\sum_{l=1}^{N-1}G(2,l). \label{eg2}
\end{eqnarray}
We can interpret $G(2,1)$ as the mean LE of all two qubit nearest
neighbors with the rest of the chain. Similar interpretations are
valid for the others $G(2,l)$.
$E_{G}^{(2)}$ is the mean of all $G(2,l)$ and it gives the
mean LE of all two qubits, independent of their distance, with the
rest of the chain \cite{scott}.
 To
define $E_{G}^{(3)}$ we 
need 
the function $G(3,l_{1},l_{2})$ with one more parameter, since now
we can have different distances among the three qubits of the
reduced state. A complete analysis of this new 
ME measure and its usefulness to detect MES is 
discussed elsewhere \cite{newmeasure}.

Table \ref{tabela1} shows the quantities given by Eqs.~(\ref{g21})
and (\ref{eg2}) for 
$GHZ_{N},EPR_{N}$, and $W_{N}$. We
note that due to translational symmetry, $G(2,1)$ and
$E_{G}^{(2)}$ are identical for 
$GHZ_{N}$ and $W_{N}$.
It is worthy of mention that depending on the value of $N$, the
states are differently classified by $G(2,1)$. A similar behavior
is observed for $E_{G}^{(2)}$ \cite{scott}. In this case, however,
$EPR_{N}$ is the most entangled state for long chains. The reason
for that lies on the definition of $E_{G}^{(2)}$. For 
$EPR_{N}$, 
$G(2,l)$ = 1 for any $l\geq2$. 
Thus, since $E_{G}^{(2)}$ is 
the average of all $G(2,l)$, for long chains $G(2,1)$ does not
contribute significantly and
$E_{G}^{(2)}\rightarrow1$. 
\begin{table}[!ht]
\vspace{-.3cm}
 \caption{\label{tabela1} Comparison among the three
paradigmatic states. }
\begin{ruledtabular}
\begin{tabular}{cccc}

 & $E_{G}^{(1)}$ & $G(2,1)$ & $E_{G}^{(2)}$ \\ \hline


 $GHZ_{N}$ & $1$ & {${2}/{3}$} & {${2}/{3}$} \\


 $EPR_{N}$ & $1$ &\large{$\frac{N-2}{2(N-1)}$} & \large{$\frac{(2N-1)(N-2)}{2(N-1)^{2}}$} \\


 $W_{N}$ & \large{$\frac{4(N-1)}{N^{2}}$} & \large{$\frac{16(N-2)}{3N^{2}}$} &

 \large{$\frac{16(N-2)}{3N^{2}}$}

\end{tabular}
\end{ruledtabular}
\end{table}

 It is worth noticing that even at the thermodynamical limit,
 $N\rightarrow\infty$,
$E_{G}^{(2)}$ and $G(2,1)$ still distinguish the three states.
However, the ordering of the states is different. 
As already explained, this is due to the contribution of $G(2,l)$,
$l\geq2$, in the calculation of $E_{G}^{(2)}(EPR_{N})$.

Now we specify to the one-dimensional Ising model in a transverse
magnetic field, which is given by the following
Hamiltonian
\begin{equation}
H=\lambda\sum_{i=1}^{N}\sigma_{i}^{x}\sigma_{i+1}^{x}
+\sum_{i=1}^{N}\sigma_{i}^{z},
\end{equation}
where 
$i$ represents the $i$-th qubit, 
$\lambda$ is a free parameter related to the inverse strength of
the magnetic field, and we work in the thermodynamical limit. 
We assume periodic boundary conditions:
$\sigma_{N+1}=\sigma_{1}$. As we have shown, for a system with
translational symmetry GE is nothing but LE of one spin with the
rest of the chain. We only need, then,  LE to obtain GE. For that
end we must calculate the single qubit (or single site) reduced
density matrix, which is 
obtained from the two qubits (two
sites) reduced density matrix. It is a $4\times4$ matrix and can
be written as\vspace{-.2cm}
\begin{equation}
\rho_{ij}=\text{Tr}_{\overline{ij}}[\rho]=
\frac{1}{4}\sum_{\alpha,\beta}
p_{\alpha\beta}\sigma_{i}^{\alpha}\otimes\sigma_{j}^{\beta},\vspace{-.4cm}
\end{equation}
where $\rho$ is the broken-symmetry ground state in the
thermodynamical limit and
$p_{\alpha\beta}=\text{Tr}[\sigma_{i}^{\alpha}\sigma_{j}^{\beta}\rho_{ij}]=
\langle\sigma_{i}^{\alpha}\sigma_{j}^{\beta}\rangle$.
$\text{Tr}_{\overline{ij}}$ is the partial trace over all degrees
of freedom except the spins at sites $i$ and $j$,
$\sigma_{i}^{\alpha}$ is the Pauli matrix acting on the site $i$,
$\alpha, \beta = 0, x, y, z$ where $\sigma^{0}$ is the identity
matrix, and $p_{\alpha\beta}$ is real. Therefore, all we need
are the ground state two-point correlation functions (CFs). By
symmetry arguments concerning the ground state \cite{nielsen} the
only non-zero CFs are $p_{00}$, $p_{xx}$, $p_{yy}$, $p_{zz}$,
$p_{0x}=p_{x0}$, $p_{0z}=p_{z0}$, and $p_{xz}=p_{zx}$. Due to
normalization $p_{00}=1$ and a direct calculation gives
$p_{xz}=p_{zx}=0$ for $\lambda\le 1$. On the other hand, the
Schwartz inequality necessarily  gives  $0\le |p_{xz}|\le
|\langle\sigma_i^x\rangle\langle\sigma_i^z\rangle|$, allowing thus
that the lower and upper bounds for entanglement be calculated for
$\lambda>1$. We plot the upper bound for entanglement by taking
$p_{xz}=0$. By continuity the true value for entanglement must
show a similar behavior.

\begin{figure}[!ht]\vspace{-.1cm}
\includegraphics[width=2.75in]{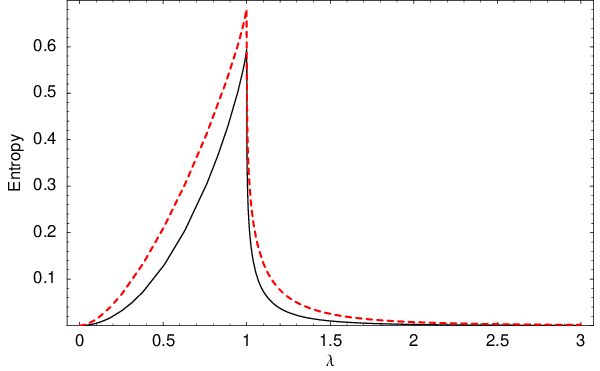}
\caption{\label{Fig_SLandSV} (Color online) Von Neumann entropy
(dashed) and GE/LE (solid) as a function of $\lambda$.}
\end{figure}

\begin{figure}[!ht]
\includegraphics[width=2.75in]{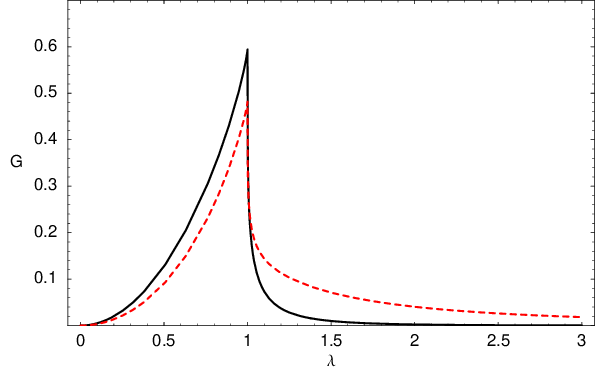}
\vspace{-0.1cm}\caption{\label{Fig G1andG21}(Color online)
$E_G^{(1)}$ (solid) and $G(2,1)$ (dashed) as a function of
$\lambda$. Both quantities are maximal at the critical point
$\lambda = 1$.}
\end{figure}

\begin{figure}[!h]
\includegraphics[width=2.75in]{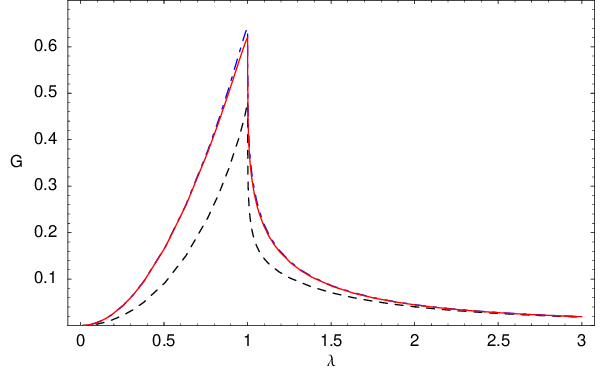}
\vspace{-0.1cm}\caption{\label{Fig_Comparacao_G2s}(Color online)
$G(2,1)$ (da\-shed\-/\-bla\-ck), $G(2,15)$ (solid/red), and
$E_G^{(2)}$ (dotted-dashed/blue) as a function of $\lambda$. We
see that $E_G^{(2)}$ is slightly different from $G(2,15)$, showing
that $G(2,l)$ saturates as $l \rightarrow \infty$.}
\end{figure}

Those CFs have been already calculated \cite{Ising original} and
we just highlight the main results. The two-point CFs and the mean
values of $\sigma^{x}$ and $\sigma^{z}$ are
\begin{eqnarray}
\langle\sigma_{1}^{x}\sigma_{l}^{x}\rangle&=&\left|\begin{array}{cccc}
g(-1) & g(-2) & \cdots & g(-l)\\
g(0) & g(-1) & \cdots & g(-l+1)\\
\vdots & \vdots & \ddots & \vdots\\
g(l-2) & g(l-3) & \cdots & g(-1)\end{array}\right|,\\
\langle\sigma_{1}^{y}\sigma_{l}^{y}\rangle&=&\left|\begin{array}{cccc}
g(1) & g(0) & \cdots & g(-l+2)\\
g(2) & g(1) & \cdots & g(-l+3)\\
\vdots & \vdots & \ddots & \vdots\\
g(l) & g(l-1) & \cdots & g(1)\end{array}\right|,
\end{eqnarray}
$ \langle\sigma_{1}^{z}\sigma_{l}^{z}\rangle =
 \langle\sigma^{z}_{1}\rangle^{2}-
 g\left(l\right)g\left(-l\right)$, $
\langle\sigma^{z}_{1}\rangle  =  g\left(0\right)$, and
$\langle\sigma^{x}_{1}\rangle=   0$ for   $\lambda\leq1$ or
$\langle\sigma^{x}_{1}\rangle= \left(1-\lambda^{-2}\right)^{1/8}$
for $\lambda>1$. Here
$g\left(l\right)=\mathcal{L}\left(l\right)+\lambda
\mathcal{L}\left(l+1\right)$, $
\mathcal{L}\left(l\right)=\frac{1}{\pi}\int_{0}^{\pi}\mathrm{d}k
\frac{\cos\left(kl\right)}{1+\lambda^{2}+2\lambda \cos (k)}$, and
$l\geq1$ is the lattice site distance between two qubits. By
tracing out one of the qubits we obtain the single qubit density
matrix, which allows us to obtain $E_G^{(1)}$ as a
function of 
$\lambda$. This is shown in Fig. \ref{Fig_SLandSV}. As a matter of
fact $E_G^{(1)}$ is maximal (with singular derivative) at the
critical point $\lambda=1$. For comparison, in Fig. 1 we plot
$S_V(\rho_j)$, which was already shown also maximal at the
critical point for the broken-symmetry state \cite{nielsen}.
We emphasize that these measures quantify entanglement in the
global system by measuring how \textit{mixed} the subsystems are.
The physical meaning behind studying ``mixedness'' lies on the
fact that the more entangled two subsystems are the more mixed
their reduced density matrix should be \cite{viola,somma}.
However, in a many-body system there are many ways in which one
could divide the global system into subsystems. The first
non-trivial generalization is to study LE of two sites with the
rest of the chain. Using 
$\rho_{ij}$ we can calculate $G(2,l)$ for the Ising model (Fig.
\ref{Fig G1andG21}).
It has a similar behavior to $E_{G}^{(1)}$, being also maximal
(with singular derivative) at the critical point. This feature
demonstrates that both a pair of nearest neighbors sites and the
sites themselves are maximally entangled to the rest of the chain
at the critical point. But this is not particular to nearest
neighbors as shown in Fig.
\ref{Fig_Comparacao_G2s}, where 
$G(2,1)$, $G(2,15)$, and
$E_G^{(2)}=\frac{1}{15}\sum_{i=1}^{15}G(2,i)$ is plotted.
$G(2,15)$ is also maximal at the critical point, 
indicating that in a 
QPT entanglement sharing at the critical point is favored by an
increase of all kind of ME.
Moreover, 
Fig. \ref{Fig_Comparacao_G2s} shows that $G(2,15)$ is only
slightly different from
$E_G^{(2)}=\frac{1}{15}\sum_{i=1}^{15}G(2,i)$. This is due to the
rapid convergence of $G(2,l)$ as $l$ is increased. At the critical
point $\lim_{l\rightarrow\infty}G(2,l)$
is $0.675$, and thus higher than the value for $GHZ_{N},EPR_{N}$,
and $W_{N}$, obtained in the thermodynamical limit, indicating
thus a genuine MES. We also remark that besides $E_G^{(1)}$,
$G(2,l)$, and
 $E_G^{(2)}$ being all maximal at the critical point, $E_G^{(1)}<E_G^{(2)}$
  for every value of $\lambda$. However
  an interesting change of ordering for $E_G^{(1)}$ and $G(2,1)$ occurs around the
  critical point.
 For $\lambda\le 1$, $E_G^{(1)}>G{(2,1)}$, but for $\lambda>1$,
 $E_G^{(1)}<G{(2,1)}$. Thence a kind of ME is
 favored in detriment of the other, depending on the system
 phase. Also, the fact that at the critical point both $E_G^{(1)}$ and
$E_G^{(2)}$ are maximal indicates entanglement sharing, such that
all the sites of the chain are strongly (quantum) correlated.
 Of course this statement is only
completely true provided that $E_G^{(m)}$ is also shown to be
maximal for any $2<m\le N-1 $ (all possible partitions).
Furthermore, the fact that $G(2,l)$ always increase as
$l\rightarrow\infty$ at the critical point suggests a kind of
diverging entanglement length. However its precise definition
demands a careful calculation of the scaling of entanglement such
as in Refs. \cite{Latorre,somma}.
 These points are left for further investigation \cite{newmeasure}.

In conclusion we have demonstrated that for an infinite Ising
chain  both $E_G^{(1)}$ and its generalization, $E_G^{(2)}$, are
maximal at the critical point. Furthermore, $E_G^{(2)}$ as defined
here is able to detect genuine ME. We remark that the behavior of
the ME measures here presented for an infinite chain is in
agreement with the Localizable Entanglement calculated for a
finite (N=14) Ising chain for the broken-symmetry state \cite{LE}.
Yet our results were obtained in a relatively simpler fashion and
could be used to infer genuine ME for systems where the
Localizable Entanglement has failed to detect QPT \cite{venuti}.
Finally, our results reinforced Osborne and Nielsen \cite{nielsen}
conjecture that at the critical point ME should be high, due to
entanglement sharing, in detriment of bipartite entanglement.

%

\begin{acknowledgments}
We thank A.O. Caldeira, E. Miranda and J. A. Hoyos for clarifying
discussions about ME and QPT. We acknowledge support from
FAEPEX-UNICAMP, CNPq, and FAPESP.
\end{acknowledgments}


\begin{thebibliography}{50}
\bibitem{footnote2} In Ref.
\cite{Nature} the authors also realized that the derivative of the
concurrence as a functions of the magnetic field diverges at the
critical point.
\end{thebibliography}
\end{document}